# *Ionic Channels in Biological Membranes*
## *Electrostatic Analysis of a Natural Nanotube*

Bob Eisenberg
Dept. of Molecular Biophysics and Physiology
Rush Medical Center
1750 West Harrison Street
Chicago Illinois 60612
USA

*Submitted, by invitation, to <u>Contemporary Physics</u>*

*Available on arXiv at*

***<u>https://arxiv.org/abs/1610.04123</u>***

*October 14, 2016*

*03:38*

# *Abstract*

Ionic channels are proteins with holes down their middle that control access to biological cells and thus govern an enormous range of biological functions important in health and disease. A substantial fraction of the drugs used in clinical medicine act directly or indirectly on channels. Channels have a simple well defined structure, and the fundamental mechanism of ionic motion is known to be electrodiffusion. The current through individual channel molecules can easily be measured, and is in fact measured in hundreds if not thousands of laboratories everyday. Thus, ionic channels are ideal objects for physical investigation: on the one hand, they are well defined structures following simple physics, on the other hand they are of general biological importance.

A simple theory of ion permeation through a channel is presented, in which diffusion occurs according to Fick's law and drift according to Ohm's law, in the electric field determined by all the charges present. This theory accounts for permeation in the channels studied to date in a wide range of solutions. Interestingly, the theory works because the shape of the electric field is a sensitive function of experimental conditions, e.g., ion concentration. Rate constants for flux are sensitive functions of ionic concentration because the fixed charge of the channel protein is shielded by the ions in an near it. Such shielding effects are not included in traditional theories of ionic channels, or other proteins, for that matter.

**_Introduction._** Ionic channels are hollow proteins with pores down their middle, found in nearly all membranes of biological cells [2, 162]. Channel proteins perforate otherwise insulating membranes and so act as holes in the walls of cells. The movement of ions (chiefly, $Na^+$, $K^+$, $Ca^{++}$ and $Cl^\Gamma$) through these channels carries the electrical charge that produces most of the electrical properties of cells and tissues. Electrons rarely carry charge more than a few angstroms in biological systems.

Ionic channels control access to the interior of cells. They are gatekeepers that govern functions of considerable biological and medical importance. Channels produce electrical signals in the nervous system; they coordinate muscle contraction, including the contraction that allows heart muscle to act as a pump. Channels transport ions in the kidneys and intestine. In nearly every cell of the body, channels control transport of ions and a wide range of other functions. It is not surprising then that a substantial fraction of the drugs used in clinical medicine act directly or indirectly on channels [155].

The physics of channels is nearly as simple as their structure. When open, channels conduct significant quantities of ions through a hole some 0.7 nm in diameter and 1!2 nm long. Channels function on the biological time scale ($>10^{!4}$ sec), which is very slow compared to the time scale of interatomic collisions ($10^{!15}$ sec) or correlated motions of water molecules ($\sim 10^{!11}$ sec). Thus, the biophysics of channels arises from only the slowest, most averaged properties of a simple physical process, diffusion, occurring in one of the simplest geometries, a 'hole in the wall' that forms a natural nanotube.

If there is any biological system of significance that can be understood as a physical system, it should be an open channel. As physical scientists, we are indeed fortunate that so simple a structure is so important biologically and medically and thus is worthy of our efforts. Too often, the biological systems that have well defined structures and so are attractive for physical analysis are rather specialized and have limited general biological significance. Not so, with channels.

**_Gating_**. Channels open and close in a stochastic process called gating and the statistics of this gating process—e.g., the mean time or probability that a channel is open—are controlled biologically to perform many important physiological functions. Most of the experimental work on channels is devoted to the discovery and description of these gating processes, and it is hard to exaggerate their biological and medical importance [90, 134, 155]. Nonetheless, gating is not a promising subject for physical analysis, at least in my opinion, until the basic structures and mechanisms involved have been discovered. They have not been yet, and I personally do not have the courage to investigate too thoroughly a mechanism we can only guess [73]. Finer scientists than I have often guessed wrong in similar circumstances in the past, and so most of my work concerns the simpler, better defined, albeit less important problem of the open channel itself, until the three dimensional structure of a channel (that has typical gating properties) is available to give clues to the underlying mechanism (see [49]).

**_Biology of Channels._** Channels come in many distinct types because they are designed and built by evolution, that is to say, by mutation and selection. The diversity of life and the molecules that do its work is one of biology's most striking characteristics. Evolution proceeds by mutation of genes, which form the blueprint of life, and the selection of those gene products that create beneficial adaptations. Beneficial adaptations increase the number of offspring of the owner of the gene and so, over time, the beneficial adaptation appears in a larger and larger fraction of the population, until it becomes 'the wild type', the typical form. Mutation and selection generate a chaotic process, which is stochastic as well because it is repeatedly reset, at random intervals, to new initial conditions, by geophysical or cosmic catastrophes.

Because genes can only make proteins, and mutations of genes are usually more or less independent events, one mutation is usually much more probable than a set of mutations. Thus, it is not surprising that evolution makes its adaptations and modifies its machines by making a single new protein (whenever it can), rather than by making a set of proteins.

Where a human engineer might build a new system to create a new function, evolution often leaves the proteins of an old system alone, and creates a new function by linking a new protein to the old system, probably for the same reason that old shared files are best left on computers. It should be no surprise then that living systems contain a staggering diversity of structures and proteins, each resulting from the concatenation of a new protein to an old structure [74, 75]. Channel proteins are no exception. Hundreds of types of channels have been discovered in the 18 years that channology has been a molecular science [39-41]. Hundreds or thousands of types remain to be discovered, I imagine.

Each type of channel has its own characteristics, but they all function by the same physical principles. We will test the working hypothesis that current flow through open channels can be understood as the electrodiffusion of ions in a charged nanotube.

More specifically, we will analyze the ionic currents that flow through open channels under more or less natural conditions, in solutions containing from 20 mM to 2 M of all types of permeant ions, when voltages are in the range $\forall 150$ mV. Fig. 1 and 2 provide idealized sketches of a channel in an experimental set-up and in a membrane. Walls 1 & 2 are insulators described by zero flux boundary conditions. End 0 and End 1 are electrodes described by inhomogeneous Dirichlet boundary conditions. The membrane (other than the channel protein itself) is in fact made of lipid with substantial surface charge. This charge is described by an inhomogeneous Neumann condition, but no ions flow through the lipid membrane away from the channel protein. The electric field exists in the lipid membrane the displacement current associated with the existence of that field is of great importance in the conduction of electrical signals in nerve and muscle fibers [168] although it is only of technological importance for measurements of single channels.

We will see how well a mean field theory of electrodiffusion [35, 53, 54] can account for these currents using these boundary conditions. In this theory, current is carried by ions moving through a charged tube of fixed structure that does not change (in the mean, on the biological time scale) with voltage, or as the concentration or type of ion is changed.

What has been striking and surprising (to those of us trained as biologists) is how much can be understood using such a spare description of electrodiffusion. Biological systems of this generality and importance often require descriptions nearly as diverse as the systems themselves, or at least that often seems to be the case. Here, a single simple description does quite well, *provided the analysis of electrodiffusion is done self-consistently*, by computing the electric field from all the charges present in the system. The variation in shape of the electric field seems to provide much of the diversity that previously could only be described, when the electric field was assumed, instead of being computed from all the charges present.

***Theory of an open channel.*** The channel protein is described in this mean field theory as a distribution of fixed charge. In the early versions of the theory—that we still use to fit experimental data quite well [32, 36, 37, 163]—the channel protein and flow of current are described by averaged one dimensional equations.

Deriving these equations from their full three dimensional form (using mathematics alone, without additional physical approximations [7, 9, 33, 35]) required us to understand a boundary condition that is scarcely described in textbooks of electricity and magnetism, even though it is the main source of the electric field for nearly any substance or molecule dissolved in water, that is to say in most things of interest to biologists and experimental chemists.

Anything that dissolves in water is likely to be an ion, or a polar molecule, as the chemists call molecules with large local but no net charge. A polar molecule has fixed charge that interacts with the fixed charge on the atoms of the water molecules. Note that polar molecules are permanently polarized, their charge is ***not*** induced polarization charge in the sense introduced by Faraday, rather they are like the electrets described in some textbooks of electricity and magnetism [76]. Water is the archetype of the polar molecule. Each of its atoms carries substantial fixed charge but these partial charges sum to zero net charge, making the water molecule neutral, overall.

The wetted surface of a protein usually has a large surface charge, determined by quantum mechanics/chemistry of the protein molecule. This surface charge is independent of ionic concentration and local electric field (for a wide range of field strengths). It does not change unless covalent bonds change; that is to say, the charge does not change unless a chemical reaction occurs. Of course, covalent bonds do change in proteins, both as metabolism occurs and when pH changes. It is in fact the change in the electric charge on proteins that make biological systems so exquisitely sensitive to pH. (A change of a few tenths of a pH unit in bodily fluids is lethal.)

Some of the surface charge of a protein is also induced by the electric field, and is traditionally described by a dielectric constant, a single number, even though the induced charge is nearly always strongly time dependent, and is often nonlinearly dependent on the electric field. Induced charge on the surface of most proteins is probably much smaller than fixed charge; it certainly is much smaller than the fixed charge lining the walls of channels. Induced charge is included (for the sake of completeness) in our original papers [7, 9] and resulting computer programs, but so far it does not seem to play an important role.

**_Interfacial surface charge_** on dissolved matter produces the electric field according to the boundary condition

$$\frac{\partial\varphi(\vec{\Gamma}_2)}{\partial n} - \frac{\partial\varphi(\vec{\Gamma}_1)}{\partial n} = -\overbrace{\frac{\sigma_0(\vec{\Gamma})}{\varepsilon_0}}^{Fixed\ Charge} - \overbrace{\frac{\sigma_2(\vec{\Gamma}_2,\varphi(\vec{\Gamma})) - \sigma_1(\vec{\Gamma}_1,\varphi(\vec{\Gamma}))}{\varepsilon_0}}^{Induced\ Charge} \qquad (1)$$

or equivalently, when induced charge is strictly proportional to the local electric field,

$$\varepsilon_{Wall}(\vec{\Gamma})\frac{\partial\varphi(\vec{\Gamma})}{\partial n} - \varepsilon_{Pore}(\vec{\Gamma})\frac{\partial\varphi(\vec{\Gamma})}{\partial n} \quad = \quad -\frac{\sigma_0(\vec{\Gamma})}{\varepsilon_0} \qquad (2)$$

Here, $\varphi(\vec{\Gamma})$ is the electric potential on the channel wall, which has a dielectric 'constant' in the range $\varepsilon_{Wall}(\vec{\Gamma}) \cong [10,30]$ compared to the dielectric coefficient $\varepsilon_{Pore} \cong [20,80]$ of the pore. The induced charge $\sigma_2(\vec{\Gamma}_2,\varphi(\vec{\Gamma}))$ is on the channel wall $\vec{\Gamma}_2$ (and depends on the local electric field, of course); the induced charge $\sigma_1(\vec{\Gamma}_1,\varphi(\vec{\Gamma}))$ is located within the pore, just next to the wall, at $\vec{\Gamma}_1$. $\varepsilon_0$ is the permittivity of free space.

**_Interfacial surface charge_** $s_0(\vec{\Gamma})$ **_is the main source of the electric field in most biological and many chemical systems_**. This fact is not widely known, unfortunately, and not properly emphasized in textbooks of electricity and magnetism, in my opinion, and has led to significant confusion among biologists, chemists, and biochemists (in particular).

Biochemists and channologists usually (if not invariably) describe the surface of a protein as a potential profile ('potential of mean force') and, forgetting that the potential of mean force is a variable output of the system, they treat the potential of mean force as a fixed input or source to the system that does not change with experimental conditions, as if it arose from an ***unchanging*** Dirichlet boundary condition. Biochemists and channologists usually (if not invariably) assume that the potential of mean force (or a rate constant derived from that potential, see eq. (14)) does not vary when the concentration of ions surrounding the protein are varied (as they often are in experiments) [84-87, 89].

In fact, in contrast to traditional assumptions, the electric field arises (mainly) from a boundary condition (i.e., eq. (1) or (2)) which becomes an unchanging Neumann condition when induced charge is negligible. If a Neumann boundary condition is imposed on a problem, the potential on the boundary will change form (as well as value) when almost any change is made in the problem. Indeed, so will the potential profile change everywhere else.

This sensitivity of systems to their boundary conditions is well known to those who have actually solved (i.e., made graphs of the solutions to) differential equations, as it is to experimental scientists who actually measure their properties, but study of boundary conditions is much less glamorous than the study of general properties of differential operators, so their significance is sometimes neglected in more general treatments of theoretical physics and mathematics.

Boundary conditions are usually important, often dominant determinants of the properties of physical systems because they describe the flow of matter, energy, and charge into the system. The biochemical case is no exception: using the incorrect unchanging Dirichlet condition is equivalent to ignoring the shielding (often called screening) of fixed charge by mobile charge. That is to say, using an unchanging potential of mean force (or rate constant) to describe the surface of a protein ignores the effect of the mobile charge (carried by ions dissolved in water) on the surface potential in the solution next to the protein.

For that reason, the traditional treatment of the surface of a protein (as an unchanging potential of mean force) is not compatible with the generally accepted treatment of ionic solutions, e.g., the Debye-Hückel, Gouy-Chapman, Poisson-Boltzmann or Mean-Spherical-Approximation (*MSA*) theories, which are in large measure an analysis of shielding and its consequences [13, 18, 42, 69, 70, 80, 127, 143, 154]. Since these theories are needed to describe the properties of ionic solutions of concentration larger than a few micro or even nanomolar, treatments of proteins that neglect shielding are unlikely to be successful in the millimolar salt solutions in which proteins are normally found. Indeed, most proteins cannot exist (in anything like normal form) in distilled water or in solutions with trace amounts of ions, suggesting that shielding plays a dominant role in the physical processes that govern the structure (i.e., folding) of proteins. It seems unfortunate that many computations (and theories) of protein folding do not include explicit ions at all. Many simulations are done with no definite concentration of ions, even though the structure of the great majority of proteins, and most of their functions (including folding) are sensitive functions of ionic concentration.

In chemical kinetics [122], biochemistry [162] or enzyme kinetics [166], rate constants are rarely if ever allowed to be a function of concentration [84-87]. *It came as a shock to realize that the usual treatment of rate constants in chemistry and biochemistry is inconsistent with the physics underlying Debye-Hückel, Gouy-Chapman, Poisson-Boltzmann and MSA treatments* and thus with the properties of the ionic solutions found in most living systems.

Unfortunately, once the potential of mean force at the surface of a protein is assumed to be independent of ionic concentration in the surrounding solution, little useful can follow, because the variation of surface potential dominates most of the equilibrium and nonequilibrium behavior of any substance dissolved in water. Nearly all equilibrium properties of proteins and of ions near the surface of proteins (e.g., the free energy per mole, i.e., the activity) are strongly affected by the concentration of other ions, that is to say by ionic strength, as has been apparent to chemists working with proteins in the lab, for at least a century.

Of course, rate constants can *sometimes* be independent of concentration of reactants, in special circumstances, for example, when the total ionic strength is held constant, while the substrate concentration is not varied enough to itself shield the fixed charge of the other reacting species or protein. Nonetheless, these are special circumstances not likely to be present in most experimental or biological systems, and they are certainly not present in open channels.

Nonequilibrium properties are at least as sensitive to shielding and the choice of boundary conditions as equilibrium properties. This is not surprising because nonequilibrium conditions: flux is often a sensitive, nearly exponential function of potential (see eq. (15)) and so rate constants (that are generally used to describe that flux) must also depend on concentration exponentially.

**PNP theory.** The flux in channels arises from gradients in the electrical potential and concentration of ions inside the channels' pore. The potential and concentration are described by two field equations whose main source are the boundary conditions (1) or (2). One field equation is the Poisson equation, that describes how the average charge produces the average potential (see eq. (3)); the other is a transport equation—in fact the Nernst Planck equation (see eq. (7))—that describes how the average potential produces flux. We call these the *PNP* equations to emphasize the importance of the Poisson equation and the relation to semiconductor physics, where the same equations are called the drift diffusion equations. The equations are coupled because flux moves charge and thereby changes the concentration and potential profile, forcing us always to deal simultaneously with both the Poisson and the drift-diffusion (i.e., Nernst-Planck) equations and solve them together.

The *PNP* equations were written and analyzed in three dimensions and the one dimensional approximation was derived, by mathematics alone, using either singular perturbation theory [9, 35] or spatial averaging [33]. Although the one dimensional version of the model seems sufficient to deal with experimental data from a wide range of channels under many conditions [32, 36, 37, 57, 130, 163]— probably because channels are so narrow and highly charged [16, 17, 20, 81]—a three dimensional theory would be more convincing, given the visual orientation of human nervous systems, and the tri-dimensionality of protein structures. A three dimensional version of the theory is computing right now (Hollerbach, Chen, Nonner, and Eisenberg, personal communication) and we hope it will be efficient enough to be useful in dealing with the mass of real experimental data.

The narrowest region of a channel is where most short range chemically specific interactions are likely to occur. In this 'selectivity filter' [90], the channel is surely one dimensional, even if the electric field is not, and so one dimensional ion transport is likely to occur. Permeation through a channel may be better represented as a one dimensional chemical reaction [57] than traditional enzyme catalysis because reactants in enzymes actually diffuse and react in a phase space of very high dimension, through an enormously complex energy landscape[59, 67]. The metaphor I once proposed ("Channels as Enzymes": [52]), may be more than the amusing tautology/oxymoron it seemed once to be.

The one dimensional theory we use to describe an open channel represents the structure of the channel's pore as a cylinder of variable cross sectional area $A(x)\,(\text{cm}^2)$ along the reaction path $x$ (cm) with dielectric coefficient $\varepsilon(x)$ and a density of charge $\rho(x)\,(\text{coul}\cong\text{cm}^{-1})$. $eN_A$ is the charge in 1 mole of elementary charges $e$, i.e. the charge in a Faraday. The charge $\rho(x)$ consists of

(1) the charge $eN_A\displaystyle\sum_k z_k C_k(x)$ of the ions (that can diffuse) in the channel, of species $k$ of charge $z_k$, and mean concentration $C_k(x)$; typically $k=\text{Na}^+,\ \text{K}^+,$ $\text{Ca}^{++}$, or $\text{Cl}^{\Gamma}$ and

(2) the permanent charge of the protein $P(x)\,(\text{mol}\cong\text{cm}^{-1})$, which is a permanent part of the atoms of the channel protein (i.e., independent of the strength of the electric field at $x$) and does not depend on the concentration of ions, etc, and so is often called the fixed charge. $P(x)$ is really quite large ($\sim0.1!1e$ per atom) for many of the atoms of a protein and wall of a channel. I imagine that the permanent charge lining the channel has an important structural role, allowing the channel=s pore to form and be stably filled with water, just as the permanent charge of a solute allows it to dissolve in water. For this reason, we should adopt the language of my friend and collaborator Wolfgang Nonner [131] and call $P(x)$ the structural charge of the channel.

(3) The dielectric charge (i.e., the induced charge which is strictly proportional to the local electric field) is not included in $\rho(x)$ because it is described by $\varepsilon(x)$. It is generally very small compared to the structural charge.

Next, we make the usual mean field assumptions that the average charge $\rho(x)$ produces an average potential $\varphi(x)$ according to Poisson=s equation and that the mean electric field $-\nabla\varphi$ captures the properties of the fluctuating electric field which are important on the slow time scale of biology. These assumptions are hardly novel; indeed, it requires some extraordinary circumstances for them not to be true, in slow highly averaged systems. *If the potential energy of mean electrical force, averaged for a msec, did not come from the mean electric charge, which source could it come from?*

Thus,

$$\varepsilon_0 \left[ \varepsilon(x) \frac{d^2\varphi}{dx^2} + \left( \frac{d\varepsilon(x)}{dx} + \varepsilon(x) \frac{d}{dx} \big[ \log_e A(x) \big] \right) \frac{d\varphi}{dx} \right] = -\rho(x) \tag{3}$$

where the average charge is given without including the small dielectric (i.e., induced) charge by the equation

$$\rho(x) \equiv eN_A \left[ P(x) + \sum_k z_k C_k(x) \right] \tag{4}$$

The boundary conditions for the potential in the real world are set by the experimental conditions: experiments, since the time of Hodgkin and Huxley [94] are best done under 'voltage clamp' conditions so that complex uncontrolled effects of voltage are avoided. Special apparatus is used to control the potentials in the baths surrounding the channel, i.e., the potential on the left is known and maintained at $V_{appl}$ and that on the right is held at zero.

$$\varphi(L) = \varphi(-\infty) = V_{applied}$$

$$\tag{5}$$

$$\varphi(R) = \varphi(+\infty) = 0$$

These boundary conditions are maintained by charge supplied to the system at its boundaries (i.e., by electrodes placed in the bath and/or inside a cell or pipette). The amount of charge necessary to maintain the potentials depends on the properties of the system, e.g., of the channels, and the experiment (i.e., whether solutions or *trans*membrane potential $V_{applied}$ are changed) and it is the need for this charge, more than anything else that makes a channel in a membrane an open system and guarantees the importance of boundary conditions in determining channel behavior.

Channels are difficult to study if the transmembrane potential $\varphi(0) - \varphi(d) = V_{applied}$ varies spontaneously in a complicated uncontrolled way. Sorting the properties of individual (types of) channels out of the mass of ionic currents flowing through all types of channels and through the interior of cells—while large capacitive currents flow through lipid membranes—is hardly ever possible (see [95] and [14] for notable exceptions). Precisely for this reason, Hodgkin, Huxley, and Katz [92], following Cole (as described in [99] and [93]), developed a feedback amplifier to supply just the current needed to control the potential to the desired value, allowing ionic channels to be studied at peace. Supplying this charge, in the resulting voltage clamped system is not easy; many biological channels are designed precisely to supply charge to change the potential, often by changing their ensemble properties in a complicated, nonlinear, time dependent way, and so designing a high quality voltage [129] or patch clamp amplifier is an interesting challenge [116].

Of course, the natural activity of membranes and channels does not occur when the voltage clamp apparatus is used. Nonetheless, natural voltage changes can easily be reconstructed by solving the Hodgkin-Huxley equations [94], which show how the current through a (voltage clamped) membrane produces the uncontrolled transmembrane potentials of a normally functioning cell. Weiss [168] is an admirable description of the classical biophysics and physiology which arose from the work of Hodgkin, Huxley, and Katz, more than anyone else. All modern systems for studying the current through one channel protein—e.g., the "patch clamp" of [151], see also [115, 116]—use the voltage clamp.

The concentrations of ions must also be controlled if the properties of channels are to be easily understood, implying the boundary conditions

$$C_k(L) = C_k(-\infty), \; C_k(R) = C_k(+\infty) \tag{6}$$

Special apparatus is not available to maintain this boundary condition, but the large volume of the baths surrounding channels, and the relatively small amounts of charge transferred through a single channel (in many cases) often guarantees that ionic flux does not significantly change ionic concentration. Such is *not* always the case, indeed such may never be the case for $Ca^{++}$ channels functioning in their normal mode, because the concentration of calcium inside cells is so small. Certainly, the absence of noticeable concentration changes must always be verified experimentally for any channel. Nonetheless, the checks are easily done and usually satisfied.

The boundary conditions (5) & (6) (at $x = \pm\infty$ of the three dimensional problem), do not map obviously and easily into boundary conditions at the ends of the channel $x = 0$, $x = d$. We have used a particular well–precedented equilibrium mapping called the built-in potential in semiconductor physics or the Donnan potential in parts of biology, see [7, 9, 33, 35]. Other treatments of the ends of the channel are under active investigation at the present time by Hollerbach, Chen, Nonner, and Eisenberg.

**Channels as Non-equilibrium Devices.** It is obvious, but nonetheless often forgotten, that a channel in a membrane, a hole in the wall, or a simulation or set of equations describing a channel or hole, describes an *open* system, a system that does *not* in itself satisfy conservation of charge, mass or energy, because charge, mass or energy must be supplied by the experimental apparatus to maintain the boundary conditions of constant concentrations and constant *trans*membrane potential. What is also not always realized is that, in the steady state—which is the case of interest here—the boundary conditions that describe a nonequilibrium and/or open system must be spatially nonuniform: if they were uniform, no charge, mass or energy could be supplied across the boundaries and the system would not be open.

An open system like a channel is a device—like a resistor—that is unlikely to be very interesting in an equilibrium state. Devices are mostly studied by engineers; it would hardly occur to them to study an amplifier, transistor, or switch 'at equilibrium'. Indeed, it is difficult to even mouth those words without smiling because it is so obvious that almost any useful property of a device or transistor involves spatial non-uniformity and thus flow.

Channels are devices, almost never at equilibrium. A channel is nearly an electrochemical wire. It *is* an electrochemical resistor, a Gaussian tube, a pillbox in an ionic solution, that only has interest when it conducts current. Like life itself, and most machines of our technology, and semiconductor devices, in particular, channels are dead at equilibrium. In that state, at equilibrium, channels show little sign of their importance or function, particularly after cremation, when they finally reach thermodynamic equilibrium, ashes to ashes, dust to dust.

Thinking about a channel, or an open system, as a device has profound consequences; energetics and thermodynamics are not emphasized (however, see [114]), input output relations (i.e., boundary conditions) are. The emphasis on boundary conditions refocuses thoughts and sometimes dramatically simplifies the analysis, by showing what should be stressed and what can be approximated. Theories and simulations that preclude flux, or that guarantee equilibrium because they use spatially uniform boundary conditions, are not useful descriptions of devices. No one studies resistors or transistors with their leads soldered together, at least not for very long, for the same reason that no one studies cadavers or their ashes if anything more lively is available.

Unfortunately, most of the simulations of molecular dynamics of channels (reviewed in [145, 148]), and most studies of their electrostatics (e.g., [104, 105, 167]), assume equilibrium and spatially uniform boundary conditions. All simulations of the molecular dynamics of proteins and most simulations of the molecular dynamics of ionic solutions make the same assumptions as far as I am aware, for example, [4, 5, 15, 22, 25, 46, 60, 78, 97, 98, 113, 119, 138, 141, 144, 148, 150, 160, 170]. Clearly, there is a vast literature of which I am unaware and I apologize to those authors (e.g., [58]) whose papers are an exception to my sweeping statement.

Whatever their exciting view of atomic detail, simulations and theories constrained to equilibrium cannot predict flow, or the properties of devices that depend on flow. Interestingly, these simulations of channels ([144-147], many further references in [148]) also do not include ions, so it is not clear how they manage to predict ionic current, e.g., [144].  Calculations of current flow in simulations of equilibrium systems, that do not include ions must, of course, give the conductance of distilled water as a result, if they are correct and converged. That conductance is many orders of magnitude less than the conductance of open channels, which the cited papers claim to predict (within a factor of two or so).

Simulations of this type can only provide an autopsy of a dead channel, even if they are done correctly including ions; they cannot predict channel function, or even glimpse its structure when it is functioning as a live wire, conducting current. Indeed, many channels and proteins cannot exist in anything like their normal state in solutions without ions (many proteins denature in distilled water into a mass rather like boiled egg white).

The effects of ions on proteins have been a well known experimental fact, crucial to the care and handling of proteins in the lab, for some hundred and fifty years. Skeptics might therefore question the relevance of simulations of proteins conducted in distilled water, without ions, or the significance of simulations of protein folding done under these conditions. Indeed, one must question whether they, like similar calculations of channels, can be converged or correct. If the protein they seek to describe cannot exist in distilled water, how can a calculation of that protein in distilled water give a valid result other than that found experimentally? How can the result be anything except a denatured protein? Or in the less extreme case, where the structure and function of the protein depends sensitively on ionic strength, how can a correct and converged simulation of the protein give a valid result when done in an indefinite concentration of ions?

Given these questions, those involved in the allocation of scientific resources might question the enormous resources that are used to simulate proteins in solutions without ions.

**_Mathematical Model_**. We turn from these general issues now to one description of flux through a channel.

The flow (i.e., the flux $J_k$ of ion $k$) through the channel is described in our mean field theory by the diffusion equation, the Nernst-Planck equation (see [18, 127]; we derive this equation below).

$$J_k = -D_k(x)A(x)\left(\frac{dC_k(x)}{dx} + \frac{C_k(x)}{RT}\frac{d}{dx}\Big[z_k F\varphi(x) + \mu_k^0(x)\Big]\right)$$

$$I = \sum_k I_k = \sum_k z_k F J_k$$

(7)

The flux $J_k$ of ions is driven by the (gradient of) concentration and electrical potential, which together form the electrochemical potential $\mu_k = RT\log_e C_k(x) + z_k F\varphi(x)$. $D_k(x)$ is the diffusion coefficient of ion $k$ in the channel's pore.

Specific chemical interactions, which cannot be easily described by the electrical or concentration terms of the electrochemical potential, can be described by an excess chemical potential, the *standard* chemical potential $\mu_k^0(x)$. This chemical term is not needed, and is in fact quantitatively insignificant in most (but not all) of the situations we have studied to date, much to our surprise: there are no shortage of reasons that water and ions in a channel should have a different standard state from water and ions in bulk solution, and I, like most workers in our field, have always assumed this would be the case. (Just consider the enormous changes in local environment that occur when an ion dehydrates and re-solvates as it enters a channel.) Nonetheless, this term $\mu_k^0(x)$ only seems important when mixtures of some ions are studied in some types of channels [29, 130, 131], probably because the fixed charge dominates the energetics of most channels [31]: even a few charges in the lining of the channel's pore produce an enormous density of fixed charge. One charge in a cylinder 6Å diameter and 10Å long is a concentration of $6 \times 10^{21}$ $cm^{-3} \approx 10$ M. Interestingly, theories like Poisson-Boltzmann are known to "become exact for large electric fields, independent of the density of hard spheres" ([p. 315 of reference 81]), "independent of interactions of molecules in the fluid phase" ([p. 972 of reference 16]. And some voltage dependent channels are thought to have as many as six charges in half that length or volume, giving ~100 M fixed charge, implying a concentration of mobile charge in the channel of about the same size. As pointed out to me by Wolfgang Nonner, such concentrations approach those of solid electrolytes: NaCl in the selectivity filter of such channel may be more like table salt than sea water, or even the water in Mono Lake (an unforgettable sight, a saturated lake, surrounded by salt columns, in Owens Valley CA, near Lee Vining).

High densities of fixed charge also help buffer the concentration of mobile charge (of the opposite sign) in the channel's pore. The important part of the channel is not exposed to the wide changes in concentration that are used in most experiments; the channel's contents are buffered by its fixed charge. Concentration independent errors in the theory can then be absorbed into its effective parameters (to some extent).

Excess chemical potentials can be analyzed (remembering that the excess chemical potential is likely to be a strong function of concentration and other variables) within the traditions of modern electrochemistry, e.g., with the mean spherical approximation of statistical mechanics [50]. Blum, Nonner, and Eisenberg are try to build a theory of the selectivity of open channels that way.

***Transport laws***. The diffusion equation used here (eq. (3)) seemed to us at first to be a crude macroscopic approximation. It turned out not to be so. Rather, eq. (3) describes the stochastic trajectories of discrete diffusing particles without much approximation.

Eisenberg, K⌒osek, and Schuss [55] derive the Nernst-Planck equation for the flux *of discrete particles* moving over a potential barrier $\varphi(x)$ of any size or shape assuming that the particles diffuse according to the Langevin equation, i.e., according to the usual laws of Brownian motion. In fact, their derivation can easily be generalized, using the methods of Schuss [124, 158] to any kind of trajectory for which conditional probabilities can be defined. Reference [55] shows that the flux can be written (here for the special case where $D_k$ is independent of $x$: the general case is given in the original paper).

$$J_k = \overbrace{C_k(L)\left(\frac{D_k}{d}\right)\text{Prob}\{R|L\}}^{\text{Unidirectional Efflux}} - \overbrace{C_k(R)\left(\frac{D_k}{d}\right)\text{Prob}\{L|R\}}^{\text{Unidirectional Infflux}} \qquad (8)$$

$$\underbrace{C_k(L)}_{\substack{\text{Source}\\\text{Concentration}}}\underbrace{\left(\frac{D_k}{d}\right)}_{\substack{\text{Diffusion}\\\text{Velocity}}}\underbrace{\text{Prob}\{R|L\}}_{\substack{\text{Conditional}\\\text{Probability}}}$$

The conditional probability $\text{Prob}\{R|L\}$ describes the probability that a trajectory starting on the *L*eft reaches an absorbing boundary on the *R*ight, when a reflecting

boundary is placed at the left, just behind the source of the trajectories (i.e., just to the left of the source).

It is traditional in chemical kinetics, e.g., [13, 142], to write equation (8) as a rate equation, namely,

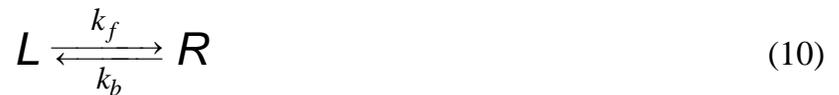

$$J_k = \overbrace{d \cdot k_f C_k(L)}^{\text{Unidirectional Efflux}} - \overbrace{d \cdot k_b C_k(R)}^{\text{Unidirectional Infflux}} \tag{9}$$

or chemical reaction

$$L \; \underset{k_b}{\overset{k_f}{\longleftrightarrow}} \; R \tag{10}$$

using "the law" of mass action. In these equations, $d$ is the channel length, and the rate constants $k_f$ and $k_b$ are defined as

$$k_f = k\{R|L\} = \frac{D_k}{d^2}\text{Prob}\{R|L\}; \qquad k_b = k\{L|R\} = \frac{D_k}{d^2}\text{Prob}\{L|R\} \tag{11}$$

In fact, reference [55] can be viewed as a stochastic derivation of the law of mass action (eq. (9)), that shows that "the law" is valid provided bath concentrations and *trans*membrane potential are maintained fixed, and $\varphi(x)$ does not vary as the concentrations $C_k(L)$ or $C_k(R)$ or *trans*membrane potential are varied. Surprisingly, if the law is valid at all, the derivation shows ***it is valid for any shape of the potential barrier***. In this way, reference [55] shows (to the considerable surprise of at least one of the authors) that the metaphor of channel permeation as a chemical reaction [52] can be exact⎯indeed it ***is*** exact when the *trans*membrane potential and bath concentrations are kept fixed in a voltage clamp experiment.

The conditional probabilities of equation (4) seem vague but in fact are precisely defined and discussed in the original publication [55]: most of the paper is devoted to their derivation and determination. The definition of the conditional probabilities must include two boundary conditions to (doubly) condition the underlying trajectories, specifying both where/how the trajectories start and where/how they end. For example, if the trajectories are those of Brownian motion, they must be described by the full, not reduced Langevin equation, to allow the double conditioning: *no matter how large the friction, the Einstein/Smoluchowski approximation of the Langevin equation cannot be used because it, being a first order differential equation, can accept only one boundary condition.* It was the use of the full Langevin equation, employing two boundary conditions that allowed us [55], to specify this problem, after many years of my frustration [43]. It was the techniques and skills of Schuss [124, 158] that provided the solution to the problem.

The conditional probabilities of equation (8)–(10) (and thus the rate constants that describe ion permeation) can be evaluated in different ways, depending on the degree of approximation of interest. If one wishes to calculate the trajectories in full atomic detail, using the techniques of molecular dynamics, the conditional probabilities can be directly evaluated from the Onsager-Machlup action formulation of Newton's laws, in the presence of thermal agitation ([133]; see Elber's modern application: [56]).

If one wishes to describe the friction in more detail, for example, as a linear process but with complex time/frequency dependence, one can use a Langevin equation with memory kernels for the friction [12, 100, 101]. If one wishes to describe the normalized friction $\beta(x)$ as a time independent process, described by a function only of position, or even by just one number independent of position, the conditional probabilities become the solutions of partial differential equations [55] as in

$$\text{Prob}\left\{R|L\right\} \equiv \frac{\int_0^\infty v p(1, v| L) dv}{\int_0^\infty v p(0, v| L) dv} \tag{12}$$

where $v$ is the velocity of the particle and the conditional probability satisfies the full forward Fokker-Planck equation [72, 158]

$$-v\frac{\partial p(x,v|L)}{\partial x} + \beta(x)\varepsilon\frac{\partial^2 p(x,v|L)}{\partial x^2} + \frac{\partial}{\partial x}\left[\beta(x)v + \frac{\partial\Phi(x)}{\partial x}\right]p(x,v|L) = 0 \qquad (13)$$

with absorbing boundary condition $p(1,v|L)=0$ for $v<0$, i.e. the boundary is absorbing *for just those trajectories leaving the system.* (If a reduced Langevin equation were used, as we, and perhaps others, tried to do for some time, hoping to avoid the mathematical complexity of the full equation, the boundary has to be absorbing for *both* arriving and leaving trajectories, which makes it somewhat difficult to put a source there, at the boundary, as is actually present in the experimental situation. Describing the experimental situation requires the full Langevin equation.) $\varepsilon$ is the normalized temperature in the customary, if unfortunate notation [158] of this field of mathematics, and is not small.

Analytical approximations are particularly neat when friction is large and simple in behavior, described by a single diffusion coefficient, a single number $D_k$ for each species $k$ of ion, using normalized units $\Phi(x) = F\varphi(x)/RT$; $V = FV_{appl}/RT$.

$$k_f = k\{R|L\} = \frac{D_k}{d^2}\text{Prob}\{R|L\} = \frac{D_k}{d^2}\cdot\frac{\exp(z_k V)}{\frac{1}{d}\int_0^d \exp z_k\Phi(\zeta)\,d\zeta}$$

$$\qquad (14)$$

$$k_b = k\{L|R\} = \frac{D_k}{d^2}\text{Prob}\{L|R\} = \frac{D_k}{d^2}\cdot\exp(z_k V)\frac{1}{\frac{1}{d}\int_0^d \exp z_k\Phi(\zeta)\,d\zeta}$$

and the current through the channel is

$$
J_k = D_k \overbrace{\frac{C_k(L)\exp\left(z_k V_{appl}\right)}{\int_0^d \exp z_k \Phi(\zeta)\, d\zeta}}^{\textit{Unidirectional Efflux}} - \; D_k \overbrace{\frac{C_k(R)}{\int_0^d \exp z_k \Phi(\zeta)\, d\zeta}}^{\textit{Unidirectional Influx}} \tag{15}
$$

These expressions can be easily generalized if $D_k$ depends on location [131].

In treating this problem, one must be quite careful in taking the limit of high friction. As is usually the case in problems involving several small parameters [108, 111, 132], the limits can be taken in different ways, and the limiting process gives different answers when they are taken in different ways (i.e., the limiting process is 'nonuniform'). Uniqueness is achieved by *taking the mathematical limit that preserves the main features of the underlying physical system and experiment.*

When taking the limit of high friction in these problems, it is essential to keep the flux fixed at a value like that observed experimentally. It is essential that the flux not be forced to zero. Otherwise, one is studying an equilibrium or nearly equilibrium system which is dead to the world we want to investigate. The driving force (i.e., the concentration and/or electrical potential gradient from one end of the channel to the other) *must not be kept fixed in the limiting process*. If the driving force were kept constant, then allowing the friction to get large forces the flux to approach zero, and such analysis must yield the equilibrium result, which can not be expected to be very useful.

A limit process with fixed driving force also forces a strictly symmetrical (in fact Maxwellian) distribution of velocities of ions. Analysis subsequent to that limit process, can only predict zero flux, if it is done in a mathematically consistent (i.e., correct) manner. One should not be surprised then if analyses or simulations that start with the reduced Langevin equation (i.e., that assume the Einstein/Smoluchowski form of the Langevin equation, without the second derivative term, which means that the distribution of velocities is strictly Maxwellian), or that explicitly assume a strictly Maxwellian distribution of velocities, or that force a strictly Maxwellian distribution of velocities (by re-adjusting the distribution of velocities, as is commonly done in molecular dynamics simulations of proteins [21, 22] and channels [145, 148] every 10–100 fsec or so) have difficulties and produce paradoxes (Ch. 7 of [3]; p. 425-430 and p. 1-75 of [118], amongst others; Ch. 8 of [4]; p. 83-92 of [78]; Ch. 6,7 & 10 of [62]; Ch. 9 & 10 of [98]).

On the other hand, if the flux is held constant while the friction goes to infinity, the distribution of velocities turns out to asymmetrical, as it must to accommodate the flux self-consistently, but the distribution is only trivially changed. It is a displaced Maxwellian [55] as has long been known in semiconductor physics [117]. The displacement is (in suitably normalized units) just the flux itself.

As simple as this displacement is, it *must* be included in any theory or calculation or simulation of molecular or Langevin dynamics of a spatially nonuniform system, a nonequilibrium system with flux across its boundaries. If, for example, a molecular dynamics simulation is done in which flux is sought, but the distribution of velocities is forced to be Maxwellian, the problem is ill-posed and no mathematical solution exists and so no solution can be found by approximate methods. Simulations cannot converge if they are in fact performed this way because the converged solution does not exist.

The simulations of nonequilibrium transport of charge carriers performed in computational electronics [82, 83, 107, 117, 140, 165] have been notably successful in analysis of experiments and prediction (i.e., design) of new results (i.e., design of new devices: see the DAMOCLES web page for a landscaped gateway to this wonderful literature [47]).

Simulations of nonequilibrium transport of charge carriers in ionic solutions seem to have been less successful (*loc. cit*). Simulations of nonequilibrium transport of charge carriers in proteins or channels seem not to have been carried out, at least as far as I know. Perhaps the difference in success between simulations of ionic solutions and semiconductors has more to do with the strikingly different methods of simulation (of the nonuniform boundary conditions, of the non-Maxwellian distribution of velocities, and of the electric field itself) than it has to do with the differences in the mechanism of charge transport in ionic solutions and semiconductors.

Of course, the simulations of semiconductor physics also treat the electrostatic field quite differently from the way it is treated in physical chemistry. The simulations of semiconductor physics use the Gummel iteration [77, 82, 103, 106, 152] to ensure self-consistent treatment of the Poisson equation (with spatially nonuniform boundary conditions) and transport laws. It is a surprising fact that the Gummel iteration seems never to have been used in simulations or analysis of charge movement in ionic solutions or proteins and that fact may have something to do with difficulties in this field as well. Perhaps, simulations of protein molecular dynamics, including folding, will be easier, will be possible with longer time steps and less numerical difficulties, and also will be more realistic, if the electric field is computed and included as it is in semiconductor calculations, using the Gummel iteration to ensure that Poisson's equation *and far field boundary conditions* are always satisfied.

Perhaps, simulations of the properties of electrolyte solutions will be more successful if they use the Gummel iteration. This is not the place to discuss this issue in detail; suffice it to say that the electric fields found transiently in ionic solutions are very much larger than $kT/e$ and so will have profound nonlinear effects that do not average to zero. These effects are not easily incorporated into theories of ionic solutions that use linearized versions of the Poisson-Boltzmann equation, no matter how sophisticated the rest of the theory is. These large electric fields are likely to produce at least some of the excluded volume effects known to occur in ionic solutions; they are also likely to affect different ionic species of the same charge (say in a solution that contains $K^+$, $Na^+$, and $Cl^-$) in similar ways producing correlated ionic movements that would appear macroscopically as flux coupling. These electric fields would be difficult to include correctly in simulations using Ewald sums and/or periodic boundary conditions. It will be interesting to see how the traditional problems of electrochemistry, that have received exhaustive attention for nearly 100 years, will respond to analysis and simulation using the Gummel iteration.

Schuss, Nadler, and Eisenberg (and others as well, no doubt) are currently trying to do a self-consistent treatment of Brownian motion, i.e., to solve Langevin and Poisson equations simultaneously and thus to understand how trajectories combine and interact to provide the atomic basis of the shielding phenomena of mean field theories.

**_Unidirectional fluxes._** In the analysis of channels, it is important to separate the flux into two components, unidirectional influx and unidirectional efflux. The total flux cannot itself be described (in any natural way) by a (single unconditional) probability, nor can the mean first passage time or contents of an ion in a channel, because a number of the unconditional quantities are infinite in perfectly finite and well posed situations, as found by [8], and explained by [55]. The flux, contents, and mean first passage times must all be replaced by the appropriate (pairs of) conditional quantities if these infinities are to be avoided.

Interestingly, for more than 50 years physiologists have directly measured these components of flux using radioactive tracers, thus correctly separating the net flux across a channel into its (*trans*) components, although as far as I know, the importance of the *cis* components of unidirectional flux was not recognized in that literature until recently ([10], see citations in that paper). The *cis* components in fact contribute a great deal to the variance of the open channel current in many conditions, because so many more trajectories enter, leave, and reenter channels than cross them, under most conditions.

The trajectories of each type of unidirectional flux can be described by other statistics besides conditional probabilities. The time an ion takes to go from $L$ to $R$ is the (conditional) first passage time $T\{R|L\}$; the number of $\{R|L\}$ trajectories within the channel is the conditional contents of the channel $C\{R|L\}$, the unidirectional flux $J\{R|L\}$ is the flux carried by the $\{R|L\}$ trajectories, and, not surprisingly, "flux equals contents over first passage time"

$$J\{R|L\} = \frac{C\{R|L\}}{T\{R|L\}} \; ; \quad J\{L|R\} = \frac{C\{L|R\}}{T\{L|R\}} \tag{16}$$

Note, however, that *total flux does **not** equal the total contents over first passage time*.

The total flux $J$ of ions is not simply related to the total contents, because the mean first passage time of all the trajectories is not well defined [8, 55] in situations like this in which there are two ways an ion can exit the channel. If even one particle starting on the left leaves on the left, then that particle takes infinite time *to reach the right* (because it never gets there), making the mean time (of all the particles) to reach the right infinite!

The mean first passage time is the sum of the first passage times of each particle divided by the number of particles. If one of those times is infinite, the sum is infinite. In this circumstance, the 'mean value operator' is a highly biased estimator—infinitely biased, in fact—of a first passage time of a trajectory *that does in fact get to the right*. Those *trans* trajectories (i.e., the ones that do in fact get to the right) are the ones we unconsciously select when we say 'mean first passage time' (having in mind the exit time or transit time) in a channel, but in fact the mathematics requires that we consider all trajectories, giving the infinite result for the unconditional mean first-passage time. When we do explicitly condition the trajectories (according to our previously unconscious and thus implicit) thoughts, we get the explicit and sensible, intuitive and finite result cited in eq. (16), but we pay the price of having to evaluate the conditioning explicitly, *cf.* eq. (12), *et seq.*

The idea that "Flux = Contents over First Passage Time" is widely, if loosely held, among physicists and engineers interested in flow (e.g., [24, 125, 126, 159]). Evidently, in problems involving two absorbing boundaries, the flux must be separated into components if it is to satisfy this equality.

***Rate Models and Transition State Theory***. The analysis of flux over barriers can be extended to derive the exponential expressions of 'Eyring' rate theory, i.e., the transition state theory widely used to describe how rate constants depend on the height of potential barriers [8, 12, 26, 38, 55, 64, 66, 72, 79, 100, 101, 110, 112, 124, 128, 136]. If the (normalized) potential profile $\Phi(x) = F\varphi(x)/RT$ is dominated by a large barrier $\Phi_{max}(x_0) = F\varphi_{max}(x_0)/RT$, and satisfies certain other criteria—e.g., if the barrier is asymptotically symmetrical and isolated from the boundary and other maxima—expressions for rate constants reduce to exponential expressions [8]. The standard expression of the Kramers' formulation of rate theory, [72, 110, 158], as found in hundreds of papers [79, 110], is recovered:

$$k_f \xrightarrow{\substack{high \\ barrier}} \underbrace{\frac{D_j}{d\sqrt{2\pi}}\sqrt{\left|z_j\Phi''(x_{max})\right|}}_{PREFACTOR} \exp\left[z_jV - z_j\Phi_{max}(x_{max})\right] \tag{17}$$

There is no controversy in the chemical literature about this expression or its prefactor. Exactly this expression is used in the extensive literature (more than 700 papers) that describe the flux over high barriers, whether the papers report simulations, theory, or experiment, whether the papers are written in the Kramers' or Eyring tradition. Indeed, Schuss, Pollak and co-workers [135-137] have shown in an elegant way how these traditions can be united rigorously: the subject of one dimensional diffusion over a barrier can be considered closed (when friction is large and simple), more or less completely reduced to known mathematics, now that the transmission factor has been evaluated by purely mathematical means [109, 137, 156, 157], and the issue of recrossings has been solved [65, 68].

The numerical value of the prefactor of eq. (17) can be estimated easily if the potential profile $\Phi(x)$ is a symmetrical parabolic barrier spanning the whole length $d$ of the channel, with maximum size $\varphi_{max}(x_{max})$, much larger than the applied (i.e., *trans*membrane) potential $V = FV_{appl}/RT$. Then, for example,

$$k_f \xrightarrow[\text{High Barrier}]{\text{Parabolic}} \underbrace{\frac{2D_j}{d^2\sqrt{\pi}} \sqrt{\left\| \frac{z_k F \varphi_{max}(x_{max})\beta}{RT} \right\|}}_{\text{PREFACTOR}} \exp\left[-z_k F \frac{\varphi_{max}(x_{max})}{RT}\right] \qquad (18)$$

These approximations (17) & (18) can be used when systems satisfy the conditions under which they were derived, e.g., when barriers are *known* to be large; isolated and asymptotically symmetrical; when the size of barriers is independent of experimental conditions. Otherwise, they should not be used, as seems obvious to physicists but seems not to be so obvious to biologists, biochemists, and (sadly) biophysicists. If a theory assuming high barriers of known fixed size is used to describe a system with low barriers, or with barriers of variable size, it is unlikely to fit the data. If such a theory is forced to fit the data, as will happen in some cases given the unhappy reality of some human behaviors, only worse will happen. Either the parameters will be forced to vary in an arbitrary way, without obvious meaning, or data showing misfits will be suppressed, or not measured at all, or even manipulated. In any case, the search for understanding will be delayed or blocked by the misuse of mathematics.

Approximate exponential expressions like (18) were originally derived and used [79, 112] before computers were available to evaluate definite integrals. Now, that they are, now that Gaussian (e.g., Gauss-Hermite, e.g. p. 217 of [161]) quadrature methods are well known that require only a handful of evaluations of the integrand, it is hardly more difficult to evaluate the exact expression than the approximate.

Further thought shows that the exact expression is less useful than it seems, in still another way, because it is likely to be used experimentally with the (sometimes tacit) assumption that barriers are *of a definite size that does not change during the experiments of interest.* In most experimental situations, however, shielding is involved in determining the barrier height, and experimental conditions change the shielding; $\varphi(x)$ and $\varphi_{max}$ are likely to change as experimental conditions are changed, e.g., as solutions or *trans*membrane potentials are changed, or drugs are applied.

Few papers using transition state theory, or rate constant models for that matter, allow for this effect, which can have large consequences because flux is usually an exponential function of $\varphi(x)$. As far as I am aware, no paper using barrier models of channels has ever calculated the height of the barrier and allowed that height to vary with ionic concentration, although in the vast literature of channels, something may easily have escaped my attention.

**_Rate Theory in Channology_**. Much of the previous discussion is made moot in the particular case of channels, because unfortunately, the version of rate theory used in channology [88, 90, 91] does not use the expressions (14) or (17),(18) found in the physical literature (e.g., [64, 79]). Rather, the traditional prefactor used in barrier theories of channels is

$$\textit{Traditional Prefactor} = \left( RT/hN_A \right) \tag{19}$$

Using the traditional prefactor produces qualitative confusion because the resulting physical meaning of the rate expression for flux is inappropriate. The traditional prefactor does not include a parameter to describe frictional interactions.

It is silly to describe flux in a pore of a protein, or an ionic solution, with an equation that does not include friction. Solutions and proteins are 'condensed phases' because they contain little empty space. Ions cannot move in such system without hitting other molecules every few femtoseconds, as was apparent some time ago [19, 23, 61]. Friction is involved in every atomic and macroscopic flux [164], on the femtosecond time scale of atomic motion [13, 123], as well as on the micro– to milliseconds biological time scale. The smaller the system the more important is friction [11, 139], and the atomic interactions that produce the friction. Proteins and channels are very small.

***Without a parameter for friction, traditional rate theory cannot predict the dependence of flux on friction found in nearly every experiment measuring current flow in condensed phases***, at least the dependence cannot be predicted in any natural way [63].

The quantitative problems produced by using the traditional prefactor are even more serious than the qualitative ones. Using the traditional prefactor produces numerical nonsense because the resulting errors in the prediction of the flux are a factor of about $2 \times 10^4$ (see Appendix of [32]; see also [43, 44]).

The word 'nonsense' in the previous paragraph may seem harsh, but the numerical error in the prefactor is so large, and has had such consequences on the history of channology, that it seems necessary. Traditional rate models found in hundreds of papers in the channel literature—following [88, 91], see [90] for references—cannot predict currents more than 0.1 pA if they use the correct prefactor. Most currents recorded from open channels are larger than 1 pA, usually much larger (say 100 pA). Thus, the model customarily used to describe open channels for some 20 years is not able to predict currents within a factor of 100 or 1,000 of those actually measured, if that model is used in the form universally accepted by physical scientists.

The traditional prefactor probably was originally used to describe enzymes and channels because of a misunderstanding of the role of entropy more than a misunderstanding of the role of friction (compare [13, 26, 112] with [88, 90, 91]): using the traditional prefactor is equivalent to ignoring the entropy change accompanying ion movement from a three dimensional bath to a (nearly) one dimensional channel (see p.1147-1157 of [13]). In any case, whatever the historical cause of the misunderstanding, once the incorrect functional form, with incorrect physical meaning and wildly incorrect numerical value is used for the prefactor, the real issues in ion permeation, the physical, chemical and biological basis of channel function, are hidden and cannot even be addressed. Little worthwhile can follow such qualitative and quantitative errors.

***Rate Constants and Barrier Heights Depend on Concentration***. Another more subtle difficulty arises when permeation, or other processes involving ions or charged species, is described by an equation (like eq. (9) or chemical reaction (like eq. (10) involving rate constants. Rate constants are historically defined as independent of concentration. The entire concentration dependence of flux is supposed to be described explicitly by the concentration variables in eq. (9); in a way, that is what the law of mass action means. We have already seen the tension between this idea, and the ideas of shielding embedded in the Gouy-Chapman/Debye-Hückel/Poisson-Boltzmann treatments of potential barriers. Those treatments imply that concentration changes potential barriers and thus rate constants.

Potential barriers have profound effects on rate constants. The rate of a reaction clearly must depend on the height of the potential barrier (*loc. cit.*) over which reactants move, indeed, we have seen that it depends exponentially. ***Thus, any variable that changes the barrier height must change the rate constant.*** In particular, if the potential barrier arises from the structural charge of a protein (as it does, at least in large measure, in channels, proteins, and enzymes, as well as a wide range of other chemical reactions), anything that shields the structural charge of the protein will modify the potential barrier and thus the flux.

Changing the concentration of reactant, or of other charged species, will certainly change the shielding of fixed charge. Indeed, in the usual view of ionic solutions, captured in the Debye-Hückel, Gouy-Chapman, and Poisson Boltzmann theories [13, 18, 27, 69-71, 80, 96, 120, 121, 127, 143], shielding phenomena are dominant determinants of nearly all the properties of ionic solutions, and changes in concentration of mobile ions is the usual way shielding is changed.

Thus, one must inevitably conclude that *changing the concentration of reactants or other charged species will change the potential barrier across which those reactants move*. The rate constant must then change. In other words, ***the rate constant must be a function of concentration***. The consequences of this statement are large because the rate constant has never been allowed to vary with concentration in the rate models of channology or biochemistry I know of; indeed, it is not usually allowed to vary with concentration in rate models in physical chemistry.

Unfortunately, approximations do not permit easy escape from this trap: the dependence of potential height on the concentration of reactant and ionic species is more likely to be exponential than weak (although of course in certain cases it can be made to be weak by using special solutions which buffer ionic strength, for example). The dependence of flux and rate constant on barrier height is in fact usually exponential. We conclude then that even when barriers are high, barrier models must re-compute the barrier height in each experimental condition of interest, because the barrier height is likely to change as experimental conditions are changed. Indeed, changing experimental conditions is likely to have an exponential effect on rate constants.

This issue is stressed here not because I enjoy being awkward or critical, but because it is a key to understanding channels, at least in my opinion, and it may be a key to understanding other proteins and biological systems as well. The success of the *PNP* equations in fitting large amounts of data measured over wide ranges of ionic strength and *trans*membrane potentials occurs precisely because the potential profiles in that theory vary widely as concentration and *trans*membrane potentials are varied in typical experiments.

We must conclude then that traditional rate models of ion movement through channels suffer from several problems: they use the wrong formulas for flux; they assume barriers independent of concentration; they assume high barriers where none need exist. It is no wonder, then, that barrier models are unable to fit experimental data taken under a wide range of conditions.

**_Rate Theory in Biophysical Chemistry._** It seems likely that similar errors occur in many areas of biophysical chemistry and molecular biology besides channology: transition states are analyzed with the incorrect prefactor in leading texts, e.g., see p. 188 of [162], and friction is never mentioned, even in qualitative discussions (see

p. 131-132 of [2], p. 38-39 of [48]), as far as I can tell. Since the effects of ion concentration on barrier height are also absent from these discussions, it seems likely that some revision in traditional enzyme kinetics will be necessary if the widely held goal of molecular and structural biology—of linking enzyme structure and function—is to be reached.

**_PNP equations_**. We try to avoid these difficulties by using the combination of eq. (3) and (15) to describe the current through open channels. In this way, we combine a description of the trajectories of diffusing particles with a self-consistent calculation of the mean electric field. We call these the Poisson-Nernst-Planck or *PNP* equations to emphasize the importance of the electric field, although they have been known in physics as the drift diffusion equations for some time.

The *PNP* equations are deceptively simple both in their physics and in their form. Physically, they are mean field equations like those of other mean field theories and they depend on the same assumptions. But the *PNP* equations differ from many mean field theories because they explicitly and self-consistently allow flux. This is very different from Debye-Hückel (etc.) theories which are confined to equilibrium, where no flux flows.

The *PNP* equations describe the rich behavior of semiconductor devices, such as switches, amplifiers, and memory elements, for example, even though they look like ordinary differential equations with much simpler behavior. The parameters of the *PNP* equations do *not* have to be adjusted to describe a transistor behaving as each of these devices. Only the boundary conditions, **_not the differential equation or its parameters,_** need to be changed to convert the device from a linear amplifier to a logarithmic amplifier or even a nonlinear switch.

The *PNP* equations are deceptively simple in this way, giving a rich repertoire of well determined behavior from a simple pair of equations. They are deceptive in other ways as well, because they cannot be integrated by the normal numerical recipes widely available in packaged programs. Those integration schemes do not work on these equations, even approximately, for fundamental reasons that are well understood mathematically [103]. Other methods work well, however.

***Solving the PNP equations: the Gummel iteration.*** Integration of the *PNP* equations is easy if a particular method called the Gummel iteration, or its equivalent, is used. The Gummel iteration was discovered some decades ago by the semiconductor community (and was discovered in my lab independently by Duan Chen, some years later) and is a general method for producing a self-consistent solution of coupled equations closely related to the self-consistent field methods used in quantum mechanics to compute orbitals.

The Gummel iteration starts with an initial guess of the potential profile, often just a linear function of position connecting the boundary values of potential. That initial guess of the potential profile is substituted into the right hand side of an integrated version of the Nernst-Planck equation (7). This substitution determines the congruent initial guess of the concentration profile $C_j(x; \textit{initial guess})$ and that guess is substituted in the right hand side of Poisson's equation (3), which is then trivially solved. The resulting estimate of potential $\varphi(x; \textit{first iterate})$ identically satisfies the boundary conditions, as do all other estimates of the potential profile. The potential profile $\varphi(x; \textit{first iterate})$ is substituted into the integrated Nernst-Planck equation (7) and so determines a first-iterate of concentration profiles $C_j(x; \textit{first iterate})$. These two iterates are consistent with each other and the boundary conditions. The two first-iterates $\varphi(x; \textit{first iterate})$ and $C_j(x; \textit{first iterate})$ are then substituted into the right hand side of Poisson's equation (3), which is again solved, now to determine the second-iterate $\varphi(x; \textit{second iterate})$, a better approximation to the potential profile. The second-iterate of potential determines a second-iterate of concentration by equation (7); together, the two second-iterates determine the third-iterate of potential, and so on for ten iterations, (which is more than enough for good convergence in almost all cases), that take only milliseconds on a typical personal computer. Once the iteration has converged, both the concentration and potential profile are known (for that set of concentrations and *trans*membrane potential, and other parameters) and so the flux and current can be determined from the output equations (15). In this way, the *PNP* equations can be easily solved to predict the current voltage relations observed in experiments.

***Comparison with experiments.*** The *PNP* equations form a map between the structure of the channel protein, represented crudely by the function $P(x)$ and the current voltage curves measured experimentally. Different types of channels have different pores made with linings of different charge. A useful and productive working hypothesis assumes that the only difference between different types of open channels is their different distributions of fixed charge $P_i(x)$, where the subscript $i$ identifies the type of channel protein, e.g., a voltage activated $Na^+$ channel, a stretch activated channel and so on [39-41, 134, 155]. Of course, the working hypothesis cannot always be true: specific chemical interactions, not captured in this simple mean field theory, will no doubt be important in ways we do not yet understand. Nonetheless, as we write these words, the current voltage relations of some 7 types of channels in a wide range of solutions can be predicted by simple distributions of fixed charge $P_i(x)$[28, 32, 36, 37, 163]. Specific chemical interactions appear only when we consider solutions containing ions of different types [30, 131] and even then remarkably little chemical information is needed: several types of ions seem to have an excess chemical potential which can be characterized by a single number, independent of concentration and potential.

The data from the porin channels is of particular interest because the locations of the atoms of that protein are known by x-ray crystallography [45, 102, 153] and amazingly even our first analysis using *PNP* recovers the correct value of charge when a mutation is made in the protein[163].

One particular kind of channel (from cardiac muscle) has been the object of extensive experimentation. This channel also appears to be strikingly simple: a fixed charge $P_{cardiac}(x) = P_0$ independent of position, with $P_0$ equal to ~$1e$, predicts the currents measured in solutions of all the monovalent cations (i.e., Li$^+$, Na$^+$, K$^+$, Rb$^+$, Cs$^+$) from 20 mM to 2 M of one type of ion, and potentials in the range ±150 mV, assuming each ion has a different diffusion coefficient [29]. The value of the diffusion coefficients are estimated by fitting theoretical predictions to the experimental data. Typically, the diffusion coefficients are some $10 \times$ less than in free solution.

This result surprised us considerably, because it shows that the same permanent charge and structural parameters (e.g., diameter and length) can fit an enormous range of data, implying that the channel is much the same whether an ion with a diameter of around 1.4Å (Li$^+$ ) or 3.9Å (Cs$^+$) fills the channel's pore. Of course, what the data really says is that any difference in channel permeation that depends on the type or diameter of the ion can be described as a diffusion coefficient, but the naïve interpretation of the result is striking and cannot be ignored: channels may be much more rigid (as measured by the average value of their properties that determine flux on the biological time scale) than any of us have expected.

**_Selectivity: properties in mixtures of ions._** The experiments just described were performed in pure solutions of the different types of ions, e.g., 20 mM NaCl on one side of the channel with 200 mM Na$^+$ on the other, or 50mM CsCl on one side and 500 mM CsCl on the other. A more common (but complex) way to study the ability of the channel to select between ions is to make mixtures of ions and apply them to both sides of the channel, e.g., 20 mM NaCl and 20 mM CsCl on one side and 200 mM NaCl and 200 mM CsCl on the other. The ability of channels to select between ions is one of their most important and characteristic properties so experiments of this type have received much attention, with probably hundreds of papers being written in the last few years on the different selectivity of different channels under varying conditions.

The properties of channels in such mixtures can be quite complex, as can the properties of mixed solutions in the bulk [6, 143], and this is not the place to discuss them in detail. Suffice it to say that the crucial experimental property called the anomalous mole fraction effect (which is called the mixed alkali effect in synthetic crystalline channels [169]) can be easily be explained by the *PNP* model if a bit of localized chemical binding is introduced [130]. Interestingly, the properties of L-type calcium channels, which have been the subject of a considerable literature, can be explained even without chemical binding, provided the dependence of fixed charge on the pH near the charged group is included in the theory [131]. In one case, where a large data set is available, all the current voltage curves *from all mixtures* of $Li^+$, $Na^+$, $K^+$, $Rb^+$, $Cs^+$ can be explained simply by including a small excess energy for $Li^+$ and $Na^+$ which arises from dehydration/resolvation as the ion enters the channel [29]. These excess chemical potentials are independent of concentration or potential.

***Generalizations.*** The striking success of such a simple model as *PNP* perplexed us for many years, until we considered its bases, physical, chemical, mathematical, and biological.

Physically, the model succeeds because the shape of the potential profile is so variable, reflecting the different shielding of fixed charge in different conditions [54]. Chemically, the insensitivity of the properties of the channel to details of the atomic structure of the channel (at least in the channels and properties studied up to now) arises probably because of a mathematical property of the *PNP* equations. These equations show that the potential profile is described (roughly speaking) by two integrations of the fixed charge profile. The current through the channel depends in turn on the integral of the potential profile, meaning that the fixed charge profile is integrated three times before it determines the variable that is experimentally measured. This is not a proof by any means that the experimental current is an insensitive function of fixed charge, but it is a plausible physical argument showing that some or many details of the charge profile are not revealed in the experimental parameters we measure. (If the fixed charge profile has regions of opposite sign, depletion layers are likely to occur that can dominate the conductance of the channel, because their high resistance is in series with the resistance of the rest of the channel [130]. In this case, and perhaps others, details of the charge profile can make a big difference.)

Biologically, such a simple model works because the fixed charge density of channels is very large (1 fixed charge lining the wall of a selectivity filter 7Å in diameter and 10Å long must be neutralized on the average by a concentration of mobile charge in the pore of around 5 M!) implying that interactions of permeating ions with the mean field are much greater than with each other [16, 81].

**Prospects**. The prospects for future development seem promising. Working with channels, we clearly must push the *PNP* theory harder and further, seeking its limits as it is used to interpret richer and richer selectivity data, e.g., including divalent ions. Working with numerical analysts, we must develop a three dimensional version of *PNP*. Working with molecular biologists—the incredibly successful molecular anatomists of our age—we must use a three dimensional version of *PNP* to confront the reality of protein structure. Working with physical chemists, we must incorporate modern ideas of equilibrium and nonequilibrium selectivity (e.g., of the *MSA* theory) into the equations. Working with physicists, we must incorporate the Gummel iteration of computational electronics (used in simulations on femtosecond and picometer scales) into simulations of proteins and ionic solutions on similar scales. Working with physiologists, we must try to explain proteins that can transport an ion against a gradient of its own electrochemical potential, using the gradient of another ion as an energy source. Perhaps our guess [34, 53] that such mediated transporters use branched channels, much as bipolar transistors do, will become a productive hypothesis.

Finally, working with our imagination, we must seek a physically specific, and anatomically justified model of gating [73], so the entire range of properties of ionic channels can become the workground of physicists and physical chemists. Perhaps our guess [34, 53] that voltage/chemical sensitive channels use branched channels to amplify their sensitivity, much as field effect transistors do, will become a productive hypothesis.

**_Conclusion_**. It seems worthwhile, as well as necessary, to compute, rather than assume the electric field when studying channels, and probably proteins and ionic solutions as well, taking care to satisfy Poisson's equation and spatially nonuniform boundary conditions. Once the shape of the electric field is studied in biology, bio-, and electro-chemistry in this way, perhaps its role will prove as important as in semiconductors. Certainly, examining the role of the electric field will keep us busy for sometime, and hopefully will help us understand and control biological systems of immense medical and thus personal significance to us all.

# *Acknowledgement*

Exploring the implications of a corrected boundary condition have occupied and astounded Duan Chen and me now for many years. It has been a joy to share the journey with him. Many gifted collaborators and friends have helped us find the way: Victor Barcilon, John Tang, Kim Cooper, Peter Gates, Mark Ratner, Ron Elber, Danuta Rojewska, Zeev Schuss, Malgorzata Kᐧosek, Joe Jerome, Chi-Wing Shu, Steve Traynelis, Eli Barkai, Jurg Rosenbusch, Nathalie Saint, Tilman Schirmer, Raimund Dutzler, Jim Lear, Le Xu, Ashutosh Tripathy, Gerhard Meissner, Carl Gardner, Wolfgang Nonner, Dirk Gillespie, Uwe Hollerbach, and Lesser Blum, more or less in chronological order. We hope they have enjoyed traveling with us as much as we have with them.

We are fortunate that this work was supported by the NSF and DARPA (grant  N65236-98-1-5409 to BE) and the NIH (to WN).

# APPENDIX
# GUIDE TO THE LITERATURE

This paper is likely to be read by many who are not familiar with the literature of channels and so I include a brief guide to the literature, following the most welcome suggestion of one of the reviewers.

The properties of channels are described in a number of recent books, e.g., [39-41, 90, 134] and [90] is the standard introductory treatment. Most papers on ionic channels are published in Biophysical Journal, the Journal of Physiology (London), the Journal of General Physiology, and some in Neuron, and most workers in the field attend the annual meeting of the USA Biophysical Society (www.biophysics.org/biophys), where several thousand papers and/or posters are presented on channels every year.

When reading these papers on channels, it is important to determine if the measurements being made are of single channel molecules or of ensembles of channels. In the former case, properties of the open channel are easily estimated from the height of the single (i.e., open) channel currents, and the properties of gating can be estimated from the statistics of the time the channel spends in its open and closed states. When ensembles are measured, it is often difficult to distinguish gating and open channel properties. Indeed, it is often difficult to be sure how many types of channels contribute to the observed currents.

Interpreting the meaning of papers on channels requires an understanding of the standard paradigms of the field. Fortunately, [90] provides a well written, readily understood summary of these paradigms. [2] and its more recent extract [1] provide wonderful introductions to molecular biology. [149, 151] provide very well done descriptions of experimental techniques, often in enough detail that the original literature need not be consulted.

Unfortunately, the interpretations of mechanism found in many of these references depend on a misuse of Eyring rate theory, in my opinion, as I have discussed and documented above and elsewhere (e.g., [28, 43, 44, 51, 53, 54]). Thus, conclusions concerning mechanism found in these references must be viewed as conjectures, to be re-examined with models of ion permeation based on generally accepted principles of ion movement in condensed phases. I do not know how many of these mechanisms will survive such re-examination and am eager to find out.

# FIGURE CAPTIONS

Fig. 1.   A sketch of a channel in a membrane in an experimental apparatus. Note that the hole in the membrane represents both the channel protein and the pore in the middle of the channel protein, a double meaning that is found very widely in the literature.  The ionic solutions are typically composed of $Na^+$, $K^+$, $Ca^{++}$, and $Cl^-$, the main physiological ions in concentrations of some 2 mM (for  $Ca^{++}$ and $K^+$ on the outside of cells) to say 150 mM (for $Na^+$ on the outside and $K^+$ on the inside of cells). [1, 2, 90, 149, 151] provide much better views of both channels and the set-ups used to measure them. The mathematical model is specified precisely in [8, 9, 35, 55], more succinctly in [130, 131].

Fig. 2. A sketch of a channel protein.  A typical ion might be 1.5 Å in diameter and the rest of the picture is more or less to scale, with the diameter of most channels being some 2 to 9 Å at their narrowest.  [1, 2, 90, 149, 151] provide much better views of channels. [45, 49, 153] present structures of channels determined by x-ray crystallography.

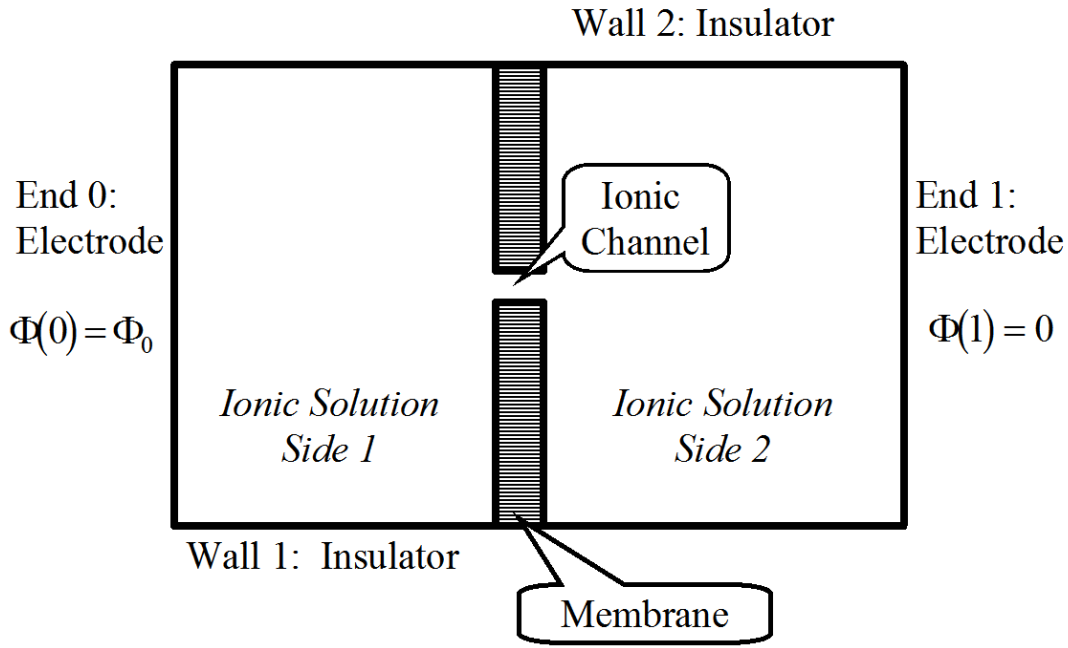

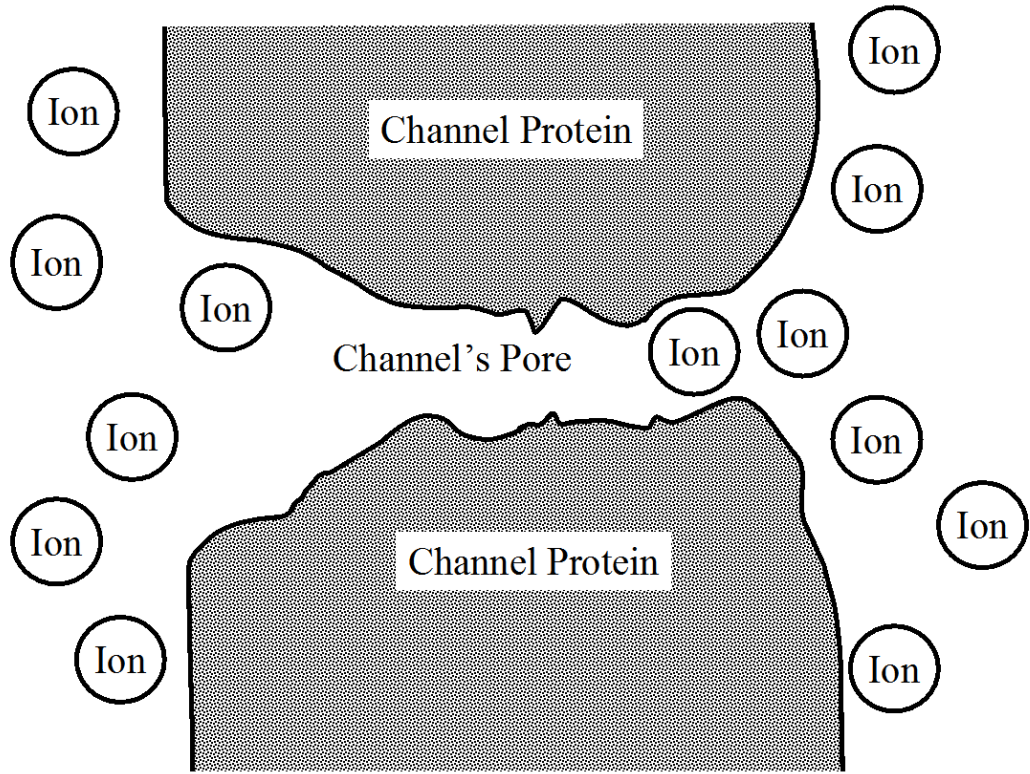